\documentstyle[amstex,amssymb,righttag,12pt]{article}
\advance\textheight\topskip

\def\pfi{\varphi}
\def\bold#1{\mbox{\boldmath{${#1}$}}}
\def\expec{\langle \ \cdot\ \rangle}
\def\state#1{\langle \, {#1}\, \rangle}
\def\del{\delta_{\alpha}}
\def\delb{{\bar \delta}_{\dot {\beta}}}
\def\vac{\vert 1 \rangle}
\def\lvac{\langle 1 \vert}
\def\s{s(\ \cdot\ )}
\def\cF{{\cal F}}
\renewcommand{\thefootnote}{\fnsymbol{footnote}}

\begin{document}

\title{Spontaneous Collapse of Supersymmetry}
\author{Detlev BUCHHOLZ$^{\dag,a}$
and Izumi OJIMA$^{\dag\dag,b}$ \\[4mm]
\it ${}^{\dag}$II.\ Institut f\"ur Theoretische Physik, 
Universit\"at Hamburg,\\ 
\it Luruper Chaussee 149, Hamburg D-22761, Germany\\[2mm] 
\it ${}^{\dag\dag}$Research Institute for Mathematical Sciences,\\
\it Kyoto University, Kyoto 606-01, Japan}

\date{}

\maketitle

\begin{abstract}
\noindent It is shown that, if generators of supersymmetry 
transformations (supercharges) can be defined in a spatially 
homogeneous physical state, then this state describes the vacuum. 
Thus, supersymmetry is broken in any thermal state and it is impossible 
to proceed from it by ``symmetrization'' to states on which an action 
of supercharges can be defined. So, unlike the familiar spontaneous 
breakdown of bosonic symmetries, there is a complete collapse of 
supersymmetry in thermal states. It is also shown that spatially 
homogeneous superthermal ensembles are never supersymmetric. \\[1mm]
\footnote[0]{${}^a$ E-mail: detlev@@x4u.desy.de}\footnote[0]
{${}^b$ E-mail: ojima@@kurims.kyoto-u.ac.jp} 
\end{abstract}

\renewcommand{\thefootnote}{\arabic{footnote}}
\setcounter{footnote}{0}
\size{12}{24pt}\selectfont

\section{Introduction}
After more than a decade of discussions, a consensus has not yet emerged on 
the fate of supersymmetry in Minkowski space 
quantum field theory at finite temperatures. There exist contradictory 
statements in the literature, ranging from the assertion that 
supersymmetry is always broken spontaneously in thermal states 
\cite{A}, through arguments in favor of supersymmetry restoration at 
sufficiently high temperatures \cite{B}, up to the claim that 
supersymmetry can be unbroken at any temperature \cite{C}. 

In this paper we reconsider the status of supersymmetry in a 
general setting including thermal states. After recalling some 
relevant facts from statistical mechanics whose significance is 
frequently ignored, we will establish the following result: If 
supercharges can be defined in a given spatially homogeneous state, 
then this state describes the vacuum. Hence supersymmetry is inevitably 
broken spontaneously in thermal states. As a matter of fact, this 
breakdown is much stronger than that of ordinary bosonic symmetries, 
where one can restore the symmetry by taking suitable averages of 
states with broken symmetry (an example being the spherical mean of a 
ferromagnetic state). In sharp contrast, such symmetrized thermal 
states do not exist in the case of supersymmetry, and we therefore 
refer to this fact as {\it spontaneous collapse} of supersymmetry 
(see below). 

We also reconsider the notion of superthermal ensemble, described by 
a supertrace, and discuss its physical significance. It turns out that 
a spatially homogeneous supertrace cannot be supersymmetric unless it 
vanishes. Hence its behavior under supersymmetry transformations is 
known from the outset and does not provide any physically significant 
information. 

In order to set the stage for our discussion, we first list some 
sources of confusion and indicate how these difficulties can be 
resolved. The general mathematical setting will be explained in 
Sec.2.\\[2mm] 
{\it Necessity for thermodynamic limit}

\noindent For a precise test of the spontaneous breakdown of a symmetry, it 
is necessary to study the thermodynamic (infinite volume) limit of the 
system under consideration. In the literature, the corresponding states 
are frequently treated as Gibbs ensembles and thermal averages 
of fields are 
presented in the form 
\begin{equation}
\langle F \rangle = Z^{-1}\ \hbox{\rm Tr}\ \hbox{\rm exp}(-\beta H) F. 
\end{equation}
This is meaningful for systems confined in a finite volume (box). In 
the thermodynamic limit, however, this formula becomes meaningless, 
since $\hbox{\rm exp}(-\beta H)$ is then no longer a trace-class 
operator. Moreover, the thermodynamic limit may not be interchanged 
with the spatial integrations involved in the definition of charge 
operators from current densities. 

These mathematical facts are frequently ignored and have led to 
erroneous statements in the literature. This problem can be 
avoided, however, by characterizing the thermal averages as expectation 
functionals $\expec$ (called states in the following) on the field 
operators which satisfy the KMS (Kubo-Martin-Schwinger) condition 
\cite{HHW}. This property survives in the thermodynamic limit and is a 
distinctive feature of thermal equilibrium states \cite{BR}.\\[2mm] 
{\it Necessity for renormalizing symmetry generators}

\noindent Another problem, closely related to the above, is the 
following one: The definition of symmetry generators by volume 
integrals of conserved Noether currents, such as 
$H = \int d^3{\bold{x}}\ \theta_{00} (x)$ in the case of the generator 
of time translations, does not make sense in thermal states in the 
thermodynamic limit. In the given example, this is obvious if one 
considers the expectation value of $H$ in a spatially homogeneous state 
with non-vanishing energy density: Infinite thermal systems contain an 
infinite amount of energy and an energy operator $H$ can therefore not 
be defined in such states. 

However, one can still define a generator ${\hat H}$ of time 
translations in thermal equilibrium states \cite{HHW} by taking 
advantage of the 
fact that these states are mixed (not pure) and hence the 
basic fields do not form an irreducible set of operators in such 
states.{\protect \footnote{In ``thermo field dynamics'' 
\cite{MTU}, one complements the basic 
fields by a set of auxiliary fields in order to deal with an 
irreducible set of operators. We do not make use of this 
formalism here.}} One can show that there exists a (state dependent) 
operator ${\tilde \theta}_{00}(x)$, commuting with all basic fields, 
such that the formal expression 
\begin{equation}
{\hat H} = \int d^3{\bold{x}}\ \big(\theta_{00} (x) - 
{\tilde \theta}_{00} (x) \big) 
\end{equation}
can be given a precise meaning as an operator in the Hilbert space 
of the given thermal state. In commutators of ${\hat H}$ 
with the underlying fields the contribution of the tilde operator drops 
out, and hence ${\hat H}$ induces the same infinitesimal time 
translation as the ill-defined expression $H$. In a sense, the passage 
from $H$ to ${\hat H}$ can be regarded as a renormalization to cancel 
out the infinities in $H$ appearing in the thermodynamic limit. 

The existence of operators commuting with the basic fields permits, in 
the case of unbroken symmetries, the construction of well-defined 
generators as described above, but it also introduces some element of 
arbitrariness: If one adds to ${\hat H}$ any operator commuting with 
the basic fields, one still obtains a generator of time translations. 
It is of interest here that, in the case of ${\hat H}$, one can remove 
this arbitrariness in a consistent manner by demanding that ${\hat H}$ 
annihilates the vector corresponding to the thermal state. (This shows, 
incidentally, that ${\hat H}$ does not have the meaning of energy since 
the energy content of a thermal state is fluctuating.) Thus the 
argument that supersymmetry must be spontaneously broken 
because thermal states carrying non-vanishing energy are not 
annihilated by $H$ is in several respects inconclusive.\\[2mm] 
{\it Spontaneous breakdown versus spontaneous collapse} 

\noindent The infinitesimal symmetry transformations of field operators 
arising from invariance properties of some Lagrangian induce linear 
mappings $\delta$ on the space of polynomials in these fields. In the 
following, we denote these polynomials generically by $F$. If $\delta$ 
corresponds to a symmetry of bosonic type, it satisfies the Leibniz 
rule 
\begin{equation}
\delta(F_1 F_2) = \delta(F_1) \, F_2 + F_1 \, \delta(F_2), \label{deriv}
\end{equation}
in an obvious notation. The analogous relation for symmetries of 
fermionic type is given in Sec.2.

The action of $\delta$ on the polynomials $F$ is always meaningful and 
can be considered in any physical state. On the other hand, the 
question as to whether the symmetry is unbroken in the sense that $\delta$
can be represented in the form\footnote{We refrain from introducing 
infinitesimal (Grassmannian) transformation 
parameters. In the case of fermionic symmetries, one then has to 
distinguish between bosonic operators and fermionic ones, replacing in 
the latter case the commutator by an anti-commutator, cf.\ Sec.2.} 
$\delta (F)=[Q,\, F]$ 
depends on the physical situation under 
consideration. Referring to the concept of thermodynamic phases, we shall   
distinguish three significant cases: (i) pure phases with unbroken 
symmetry, (ii) pure phases with broken symmetry which  
can be restored, however, by proceeding to suitable mixed phases, and 
(iii) phases, pure or mixed, with spontaneously collapsed symmetry which
cannot be restored.  

Here the notion of pure and mixed thermodynamic phases should not be 
confused with that of pure and mixed states. A pure phase is 
characterized by sharp c-number values of macroscopic order parameters, 
which are statistically 
fluctuating in the case of mixed phases. Any thermodynamic phase, pure or 
mixed, is described by a mixed state; it is only in the case of vacuum states
that the notions of pure phase and pure state coincide \cite{BR}. 

We recall that a state $\expec$ describing a pure thermodynamic phase
has the cluster property 
(absence of long range order), i.e., it holds for any $F_1, F_2$ that 
\begin{equation}
\langle F_1({\bold{x}}) F_2 \rangle - \langle F_1({\bold{x}}) \rangle 
\langle F_2 \rangle \rightarrow 0 \qquad \mbox{as }\vert {\bold{x}} \vert 
\rightarrow \infty.
\end{equation}
Here $F({\bold{x}})$ denotes the polynomial obtained from $F$ by 
shifting the spacetime arguments of the underlying fields by 
${\bold{x}}$. If a state $\expec$ describes a statistical 
{mixture} of phases, it can be decomposed into pure phases 
$\expec_{\theta}$, 
\begin{equation}
\expec = \sum_{\theta} w_{\theta} \expec_{\theta}, 
\end{equation}
where $\theta$ is an order parameter labeling the pure phases and the 
weight factors $w_{\theta}$ are non-negative numbers which add up to 1. 
(In the case of a continuum of phases, the summation need be replaced 
by an integration with respect to a probability measure.) An important 
fact about this 
central decomposition of states is its 
uniqueness, which will be used later. For a thorough exposition of 
these facts, we refer to \cite{BR}. 

Returning now to the issue of symmetry, we consider any state $\expec$ 
describing a pure thermodynamic phase. By the reconstruction theorem 
\cite{GNS}, there exists a corresponding Hilbert space of vectors, 
describing this phase as well as all states which can be reached from 
it by the action of polynomials $F$ in the fields. As indicated above, there 
are then the following three possibilities. 

(i) There exists an 
operator $Q$ on this Hilbert space which generates the symmetry transformation 
$\delta$ as described above. The symmetry is then said to be unbroken in 
this phase. 

A frequently used test as to whether this situation is realized is 
given by: 
\begin{equation}
\langle \delta (F) \rangle = 0 \quad  \hbox{\rm for all}\ F \, ?  \label{Gld}
\end{equation}
If the answer is affirmative, one can consistently define an operator 
$Q$ with all desired properties. Yet it is sometimes overlooked that 
this test provides only a sufficient condition for the existence 
of such a $Q$. This can heuristically be understood if one thinks of a 
spatially inhomogeneous situation, e.g., a drop of liquid surrounded by 
gas. The corresponding state is then not invariant under 
translations and thus does not pass the test (\ref{Gld}) for the 
infinitesimal translations $\delta$. Nevertheless, translations can be 
defined on the underlying Hilbert space since the effect of shifting 
the drop can be described by the action of polynomials in the 
fields on the state vectors: One annihilates the constituents of the 
drop and creates them again at the shifted position. 

In the case of spatially homogeneous states such as the vacuum, one can 
sometimes show that (\ref{Gld}) provides also a necessary condition 
for the existence of $Q$. However, the arguments given to that effect in the 
literature (see, for example, \cite{PTW}) are not conclusive in the 
case of thermal states because of the 
abovementioned difficulties in the definition of generators. The 
status of the test (\ref{Gld}) thus needs a close re-examination.

(ii) The second possibility is that the symmetry is broken in 
the pure phase $\expec$, but one can proceed to a corresponding 
symmetrized mixed phase where the symmetry is restored and generators 
$Q$ can be defined. 

The familiar example of this kind already mentioned is 
the case of a ferromagnet. If $\expec_{\bold{\theta}}$ 
describes such a state with sharp direction ${\bold{\theta}}$
$(=(\vartheta, \pfi))$ of magnetization, spatial rotations cannot be 
defined on the corresponding Hilbert space (except for the rotations 
around the axis ${\bold{\theta}}$). But the mixed phase 
corresponding to the spherical average of these states, 
$\expec = \int d{\bold{\theta}}\ \expec_{\bold{\theta}}$, 
passes the test (\ref{Gld}) with respect to infinitesimal rotations~$\delta$. 
Hence, there exist generators for rotations on the 
corresponding ``enlarged'' Hilbert space. In other words, while the 
result of a rotation cannot be described in the Hilbert space of each 
pure phase since it cannot be accomplished by the action of  
polynomials in the local fields on the corresponding vectors, it is still 
meaningful to speak about the action of rotations on these states. 
Generators inducing this action can be defined in the Hilbert space of 
the symmetrized state $\expec$ since it comprises states with arbitrary 
directions of magnetization. 

The situation described in this example illustrates the spontaneous 
breakdown of a symmetry: The symmetry is broken in a pure thermodynamic 
phase but restored in a suitable mixture 
where one can define corresponding generators. This situation 
prevails in the case of bosonic symmetries described by a (locally) 
compact group. It is this case which is usually taken for granted. 

(iii) There is, however, a third possibility which is of 
relevance to the 
case of supersymmetries and which, to the best of our knowledge, has 
not been discussed so far in the literature. Namely, it may happen that 
a symmetry is broken in some pure phase, but there is no 
corresponding ``symmetrized'' 
mixed phase such that an action of generators $Q$ of 
the symmetry can be defined in it. Thus, whereas the symmetry 
transformations 
$\delta$ of the fields are still well defined, the idea of transformed 
state vectors becomes meaningless. We call this case 
{\it spontaneous collapse} of symmetry. 

In view of the points raised, a thorough discussion of the fate of 
supersymmetries in thermal states seems desirable. Since it 
requires a general mathematical 
setting which may not be so well known, we recall in the 
first part of the subsequent Sec.2 some relevant mathematical notions 
and facts, and then turn to the analysis of supercharges. In Sec.3 we 
discuss the role of superthermal ensembles and of supertraces. The 
paper concludes with a brief discussion of the physical significance of 
our results. 

\section{Status of Supercharges}
The assumption that a quantum field theory is supersymmetric implies 
that there exist Lorentz-covariant anti-local spinorial currents 
\begin{equation}
j_{\mu \alpha}(x), \qquad j_{\nu {\dot {\beta}}}^{\dag}(x)
\end{equation}
which are the hermitian conjugates of each other and are conserved,  
\begin{equation}
\partial^{\mu} j_{\mu \alpha}(x)= \partial^{\nu} 
j_{\nu {\dot {\beta}}}^{\dag}(x)=0.
\label{conserv}
\end{equation}
As is well known, the singular nature of field and current operators 
at a point, generically denoted by $\pfi(x)$ (with tensor and spinor 
indices omitted), requires us to smear them with test functions $f$, i.e.,
smooth functions on ${\Bbb R}^4$ with compact support, 
\begin{equation}
\pfi(f) = \int d^4x\ \pfi(x) f(x). \nonumber
\end{equation}
In the following, we use the notation $F$ for polynomials in smeared fields
and currents, 
\begin{equation}
F = \sum c \, \pfi(f_1) \, \pfi(f_2) \cdots \pfi(f_n), 
\end{equation}
and denote by 
${\cal F}$ the set of these polynomials (forming an algebra). The 
space-time translations act on $F \in {\cal F}$ by 
\begin{equation}
F \mapsto F(x) = \sum c \, \pfi(f_{1,x}) \, \pfi(f_{2,x}) \cdots 
\pfi(f_{n,x}), \quad x \in {\Bbb R}^4
\end{equation}
where $f_x$ is obtained from $f$ by setting $f_x(y) = f(y-x), \ 
y \in {\Bbb R}^4$. Making a choice of Lorentz frame, we write 
$x = (x_0,\ \bold{x})$ and use also the shorthand notation for 
spatial and temporal translates $F({\bold{x}}) = F(0, {\bold{x}}), 
F(x_0) = F(x_0, {\bold{0}})$. We also introduce the notation 
${\cal F}_{\pm}$ for bosonic/fermionic operators, i.e., polynomials in 
the smeared field operators containing an even/odd number of fermionic 
fields in each monomial. Then, we can define the supersymmetry 
transformations by 
\begin{equation}
\del(F_{\pm}) = \lim_{R \rightarrow \infty} \int d^4x\ 
g(x_0)h(\bold{x}/R)\ [j_{0 \alpha}(x), F_{\pm}]_{\mp} \label{super}
\end{equation}
and analogously $\delb$ in terms of $j_{\nu {\dot {\beta}}}^{\dag}(x)$, 
where $F_{\pm} \in {\cal F}_{\pm}$. Here real test functions $g \in 
{\cal D}({\Bbb R})$ and $h \in {\cal D}({\Bbb R}^3)$ are so chosen that 
$\int dx_0\ g(x_0)=1, h(\bold{x})= 1$ for $\vert \bold{x} \vert \leq 1$.
{}From current conservation, Eq.(\ref{conserv}), and local (anti-)
commutativity, it follows that $\del$ and $\delb$ exist as linear maps 
acting on ${\cal F}={\cal F}_{+} + {\cal F}_{-}$ and that they do not 
depend on the choice of $g, h$ satisfying the stated conditions. 
Moreover, for given $F_{\pm}$ the limit in Eq.(\ref{super}) is 
attained for some finite $R$, so the images $\del(F_{\pm})$ are again 
operators belonging to ${\cal F}$ in accord with the more formal 
definition of supersymmetry transformations of fields in the Lagrangian 
framework. As a matter of fact, it holds that $\del({\cal F}_{\pm}) 
\subset {\cal F}_{\mp},\ \ \delb({\cal F}_{\pm}) \subset 
{\cal F}_{\mp}$, and hence the mappings can be applied to the elements 
of ${\cal F}$ an arbitrary number of times. They are anti-derivations 
satisfying the following ``graded'' Leibniz rule: 
\begin{equation}
\del (F_{\pm} F) = \del(F_{\pm}) F \pm F_{\pm} \del(F),\label{anti1}
\end{equation}
for $F_{\pm} \in {\cal F}_{\pm}, F \in {\cal F}$, and similarly for 
$\delb$. We also note their behavior under hermitian conjugation 
following from the hermiticity properties of the currents, 
\begin{equation}
\del (F_{\pm})^{\dag} = {\mp}\ 
{\bar {\delta}}_{\dot {\alpha}}(F_{\pm}^{\dag}). \label{hrmt}
\end{equation}

To put the fundamental relation of supersymmetry in a state-independent 
form, we also introduce the derivation arising from the time 
translations, 
\begin{equation}
\delta_0(F)=-i{d \over dx_0}F(x_0) \vert_{x_0=0}. 
\end{equation}
Note that the derivation $\delta_0$ and the anti-derivations $\del, 
\delb$ commute with the spatial translations, i.e., it holds that 
$\delta(F({\bold{x}}))=(\delta(F))({\bold{x}})$. 

The fundamental 
relation of supersymmetry can now be expressed as follows: 
\begin{equation}
{\bar \delta}_{\dot 1} \circ \delta_1 + \delta_1 \circ 
{\bar \delta}_{\dot 1} + {\bar \delta}_{\dot 2} \circ \delta_2 + \delta_2 
\circ {\bar \delta}_{\dot 2} = 4 \delta_0, \label{susy}
\end{equation}
where $\circ$ denotes the composition of the respective maps on 
${\cal F}$. 
This relation of maps is meaningful independently of the specific 
choice of a state. It follows either from the very definition of 
supersymmetry transformations of fields or can be verified in any 
state where supercharges can be defined as generators of supersymmetry 
transformations, for instance, the vacuum. The crucial point is that, 
because of the fermionic nature of supercharges, cancellations take 
place in the ``mixed terms'' where supercharges stand to the left and 
right of field operators. 

We consider now any state $\expec$ on ${\cal F}$ which is invariant 
under spatial translations, namely, we assume that $\expec$ has 
the properties $\langle c_1 F_1 + c_2 F_2 \rangle = c_1 
\langle F_1 \rangle + c_2 \langle F_2 \rangle$ (linearity), \ 
$\langle F^{\dag}F \rangle \geq 0$ (positivity), \ 
$\langle {\bold{1}} \rangle =1$ (normalization) and 
$\langle F({\bold{x}}) \rangle = \langle F \rangle$ (invariance). It is 
our aim to show that $\expec$ must be the vacuum if supersymmetry is 
not broken in this state. The argument will be given in several steps. 

\noindent (a) By the reconstruction theorem \cite{GNS}, there exists a 
Hilbert space ${\frak H}$ and a distinguished unit vector $\vac$ such that 
the set of vectors $F\, \vac,\ F \in {\cal F}$, is dense in ${\frak H}$ 
and 
\begin{equation}
\langle F \rangle = \lvac\, F\, \vac
\end{equation}
holds for any $F \in {\cal F}$.\footnote{Strictly speaking, one should 
distinguish between the ``abstract'' elements $F \in {\cal F}$ and 
their concrete realization as operators on ${\frak H}$ which depends on 
the given state $\expec$. Since there is no danger of confusion, we use the 
present simplified notation.} First, we show that the Bose-Fermi 
superselection rule is not broken spontaneously in any such state. Let 
$F_{-} \in {\cal F}_{-}$ be a fermionic operator. 
{}From our assumption of the invariance of $\expec$ under 
spatial translations it follows that $\langle F_{-} \rangle = 
\langle F_{-}({\bold{x}}) \rangle = \lvac\, F_{-}({\bold{x}})\, \vac = 
(1/{{\vert} V {\vert}})\int_V d^3{\bold{x}}\ \lvac\, F_{-}({\bold{x}})\, 
\vac$, where $V$ denotes any bounded spatial region in ${\Bbb R}^3$ 
and ${\vert} V {\vert}$ its volume. In the limit of $V \nearrow 
{\Bbb R}^3$ the right-hand side  of this equality can be shown to 
vanish because of the following bound on the norm of the spatial mean 
of fermionic vectors, 
\begin{eqnarray}
& &\Vert {1 \over {\vert V \vert}} \int_V d^3{\bold{x}} F_{-}({\bold{x}})\, 
\vac \Vert^2 
+ \Vert {1 \over {\vert V \vert}} \int_V d^3{\bold{x}} 
F_{-}({\bold{x}})^{\dag}\, \vac \Vert^2  \nonumber \\
&=&{1 \over {\vert V \vert}} \int_V d^3{\bold{x}} \, 
{1 \over {\vert V \vert}} \int_V d^3{\bold{y}} \ 
\langle\, [\, F_{-}({\bold{x}})^{\dag},\, F_{-}({\bold{y}})\, ]_+ \,\rangle 
\nonumber \\
&\leq& {1 \over {\vert V \vert}} \int d^3{\bold{z}}\ \vert\, \langle\, 
[\, F_{-}({\bold{z}})^{\dag}, F_{-}\, ]_+ \,\rangle\, \vert,  \label{cl}
\end{eqnarray}
where we made use of the invariance of $\expec$ under spatial 
translations. Note that the latter integral exists since the 
anti-commutator vanishes for large spatial translations~${\bold{z}}$. 
{}From this, we conclude that 
\begin{equation}
\langle F_{-} \rangle = 0 \quad \hbox{\rm for}\  F_{-} \in {\cal F}_{-} 
\label{B-F}
\end{equation}
which asserts the validity of the Bose-Fermi superselection rule. 

\noindent (b) We say that supersymmetry is {\it implementable in the state} 
$\expec$ if there exist operators $Q_{\alpha}$ and 
$Q_{\dot {\beta}}^{\dag}$ (hermitian conjugate of $Q_{\beta}$) which have the 
vectors $F\, \vac, F \in {\cal F}$, in their domains of definition and 
satisfy 
\begin{eqnarray}
Q_{\alpha} F_{\pm}\, \vac &=& \del(F_{\pm})\, \vac
\pm F_{\pm} Q_{\alpha}\, \vac, \label{unbr1}\\
Q_{\dot {\beta}}^{\dag} F_{\pm}\, \vac &=& \delb(F_{\pm})\, \vac \pm 
F_{\pm} Q_{\dot {\beta}}^{\dag}\, \vac. \label{unbr2}
\end{eqnarray}

\vskip12pt

\noindent {\bf Remarks}: (i) We do not assume from the outset that 
$Q_{\alpha}\, \vac = Q_{\dot {\beta}}^{\dag}\, \vac =0$ or that 
$Q_{\alpha}$ and $Q_{\dot {\beta}}^{\dag}$ commute with translations 
because of the ambiguities involved in the definition of generators in 
the case of thermal states, cf. Sec.1. 

\noindent (ii) Picking arbitrary vectors $\vert \alpha \rangle, 
\vert {\dot {\beta}} \rangle$ in the domain of all operators in 
${\cal F}$, one can always (i.e., irrespective of the occurrence of 
spontaneous symmetry breakdown) define consistently linear operators 
${\hat Q}_{\alpha}, {\check Q}_{\dot {\beta}}$ if $\vac$ has the 
property of being separating for ${\cal F}$, i.e., if $F\, \vac = 0$ 
implies $F=0$ for $F \in {\cal F}$. (This property holds for vacuum 
states by the Reeh-Schlieder theorem \cite{GNS} and also for thermal 
equilibrium states as a consequence of the KMS condition \cite{BR}, 
cf. below). One simply puts  
\begin{eqnarray}
{\hat Q}_{\alpha} F_{\pm}\, \vac &=& \del(F_{\pm})\, \vac \pm F_{\pm} 
\vert {\alpha} \rangle, \\
{\check Q}_{\dot {\beta}} F_{\pm}\, \vac &=& 
{\bar \delta}_{\dot {\beta}}(F_{\pm})\, \vac \pm F_{\pm} \ \vert 
{\dot {\beta}} \rangle. 
\end{eqnarray}
In this formulation, spontaneous breakdown of supersymmetries means 
that, for no choice of $\vert \alpha \rangle, 
\vert {\dot {\alpha}} \rangle$, the operators ${\hat Q}_{\alpha}, 
{\check Q}_{\dot {\alpha}}$ are the hermitian conjugates of each other. 
They then 
have very pathological properties (e.g., are not closable \cite{Reeh}) 
and thus are not physically acceptable. 

\vskip12pt

We will now show that, if supersymmetry is implementable in the state 
$\expec$ in the sense specified above, then it holds that 
\begin{equation}
\langle \del (\ \cdot\ ) \rangle= \langle \delb(\ \cdot\ ) \rangle = 0, 
\label{sym}
\end{equation}
i.e., the state $\expec$ passes the familiar test for symmetry. It 
follows from the Bose-Fermi superselection rule that, for any $F_{+} 
\in {\cal F}_{+}$, 
\begin{equation}
\langle \del(F_{+}) \rangle = \langle \delb(F_{+}) \rangle = 0, 
\end{equation}
because of $\del(F_{+}), \delb(F_{+}) \in {\cal F}_{-}$. 
In order to show that these expressions vanish also for 
fermionic operators $F_{-} \in {\cal F}_{-}$, 
\begin{equation}
\langle \del(F_{-}) \rangle = \langle \delb(F_{-}) \rangle = 0,\label{unb}
\end{equation}
we make use of Eqs.(\ref{unbr1}), (\ref{unbr2}). Combining these 
relations and the commutativity between the anti-derivation $\del$ and 
spatial translations ${\bold{x}}$, we obtain, as in step (a), 
\begin{eqnarray}
&&\langle \del(F_{-}) \rangle = {1 \over {\vert V \vert}}\int_V d^3{\bold{x}}\
\lvac \, \del(F_{-}({\bold{x}}))\, \vac \nonumber \\
&&= {1 \over {\vert V \vert}}\int_V d^3{\bold{x}}\ 
\lvac\, (Q_{\alpha} F_{-}({\bold{x}}) 
+ F_{-}({\bold{x}}) Q_{\alpha})\, \vac. 
\end{eqnarray}
Hence, by making use of the Cauchy-Schwarz inequality, we arrive at the 
estimate 
\begin{equation}
\vert \langle \del(F_{-}) \rangle \vert \leq \Vert Q_{\alpha}^{\dag}\,  
\vac \Vert \cdot \Vert {1 \over {\vert V \vert}} \int_V d^3x\ 
F_{-}({\bold{x}})\, \vac \Vert +  \Vert {1 \over {\vert V \vert}} 
\int_V d^3x\ {F_{-} ({\bold{x}})}^{\dag}\, 
\vac \Vert \cdot \Vert Q_{\alpha}\, \vac \Vert
\end{equation}
for arbitrary $V$. The right-hand side of this inequality vanishes for 
$V \nearrow {\Bbb R}^3$ according to relation (\ref{cl}) which shows 
that $\langle \del(F_{-}) \rangle = 0$ for $F_{-} \in {\cal F}_{-}$. 
By the same token we obtain also $\langle \delb(F_{-}) \rangle = 0, \ 
F_{-} \in {\cal F}_{-}$, which completes the proof of relation (\ref{unb}). 
As a consequence of the fundamental relation Eq.(\ref{susy}) 
characterizing supersymmetry, the invariance of $\expec$ under time 
translations automatically follows:
\begin{equation}
\langle \delta_0(\ \cdot\ ) \rangle =0. \label{inv}
\end{equation}

We emphasize that, in the above discussion, we did not 
assume the cluster property with respect to spatial translations, and 
hence, the result is valid even if the state $\expec$ describes an 
arbitrary mixed thermodynamic phase. The only condition on $\expec$ is 
that supercharges can be defined. 

\vskip12pt

\noindent (c) In the next step, we show that, if a state $\expec$ is 
supersymmetric in the sense of equation (\ref{sym}) and complies with 
the Bose-Fermi superselection rule (\ref{B-F}), then it is a vacuum 
state. More precisely, it is invariant under space and time 
translations and there are corresponding generators satisfying the 
relativistic spectrum condition (positivity of energy in all Lorentz 
frames). To prove this statement, we consider the expectation values  
$\langle F^{\dag} \delta_0 (F) \rangle$ for $F \in {\cal F}$. Since 
$F=F_{+} + F_{-}$ with $F_{\pm} \in {\cal F}_{\pm}$ and 
$\langle F_{\pm}^{\dag} \delta_0(F_{\mp}) \rangle =0$ by 
(\ref{B-F}), we may restrict our attention to the expectation values 
$\langle F_{\pm}^{\dag} \delta_0(F_{\pm}) \rangle$. According to the 
fundamental relation of supersymmetry, it holds that 
\begin{equation}
4 \langle F_{+}^{\dag} \, \delta_0(F_{+}) \rangle= \langle F_{+}^{\dag} \,
({\bar \delta}_{\dot 1} \circ {\delta}_1 + {\delta}_1 \circ 
{\bar \delta}_{\dot 1} + {\bar \delta}_{\dot 2} \circ {\delta}_2 + 
{\delta}_2 \circ {\bar \delta}_{\dot 2})(F_{+}) \rangle, 
\end{equation}
and we take a close look at the terms appearing on the right-hand side. 
Making use of the fact that the $\del, \delb$ are anti-derivations 
and of Eq.(\ref{hrmt}), we have 
\begin{eqnarray}
{\bar \delta}_{\dot 1}(F_{+}^{\dag} \, 
{\delta}_1(F_{+}))&=& {\bar \delta}_{\dot 1}(F_{+}^{\dag}) \,
{\delta}_1(F_{+}) + F_{+}^{\dag} \, 
{\bar \delta}_{\dot 1} ({\delta}_1(F_{+})) \nonumber \\
&=& -{\delta}_1(F_{+})^{\dag} \, 
{\delta}_1(F_{+}) + F_{+}^{\dag} \, 
{\bar \delta}_{\dot 1} \circ {\delta}_1(F_{+}).
\end{eqnarray}
Since $\langle \, {\bar \delta}_{\dot 1}( \ \cdot \ )\, \rangle =0$, 
therefore, we find that 
\begin{equation}
\langle F_{+}^{\dag} \, 
{\bar \delta}_{\dot 1} \circ {\delta}_1(F_{+}) \rangle = 
\langle {\delta}_1(F_{+})^{\dag} \, {\delta}_1(F_{+}) \rangle  \geq 0
\end{equation}
and a similar argument applies to the remaining terms. So 
$\langle F_{+}^{\dag} \, {\delta}_0(F_{+}) \rangle \geq 0$
and the same result holds if one replaces $F_{+}$ by 
$F_{-} \in {\cal F}_{-}$. Putting together all this, we arrive at 
\begin{equation}
\langle F^{\dag} \, {\delta}_0(F) \rangle \geq 0 \qquad 
\hbox{\rm for }F \in {\cal F}. 
\end{equation}

Now, since ${\delta}_0$ is a derivation satisfying the Leibniz rule 
(\ref{deriv}), ${\delta}_0(F)^{\dag}= - {\delta}_0( F^{\dag} )$ for 
$F \in {\cal F}$, and $\langle {\delta}_0(\ \cdot\ ) \rangle=0$, the 
operator $P_0$ given by 
\begin{equation}
P_0 F \, \vac = \delta_0(F) \, \vac \qquad \hbox{\rm for }F \in {\cal F}
\end{equation}
is well defined and hermitian\footnote{Making use of temperedness of 
the underlying fields, one can show that $P_0$ is even essentially 
self-adjoint on its domain of definition.} and satisfies 
$P_0\, \vac= \delta_0 ({\bold 1})  \vac = 0$. 
{}From the lower bound 
\begin{equation}
\lvac \, F^{\dag}\, P_0\, F \, \vac = \langle F^{\dag}\delta_0(F) \rangle 
\geq 0, 
\end{equation}
it follows that $P_0$ is a positive operator, so, in view of $P_0\, \vac=0$, 
we conclude that $\vac$ is a {ground state} for $P_0$. 

To show that the state $\expec$ is a ground state in any Lorentz 
frame, we make use of the spinorial transformation properties of the 
supercurrents. They imply that the fundamental relation (\ref{susy}) 
holds for the transformed maps ${\del}', {{\delb}}{'}$ and ${{\delta}_0}'$ 
in any Lorentz frame. Moreover, since a change of Lorentz frame amounts 
to a linear transformation of these maps, i.e., 
\begin{equation}
{\del}'= A_{\alpha}{}^{\beta} {\delta}_{\beta}, \ \ \ 
{{\bar {\delta}}_{\dot {\alpha}}}{'} = 
{\overline {A_{\alpha}{}^{\beta}}}\, 
{\delb}={{\bar A}_{\dot \alpha}{}^{\dot \beta}} \delb
\end{equation}
with $A \in SL(2,\, {\Bbb C})$, it follows that, if supersymmetry is 
unbroken in some Lorentz frame in the sense of 
Eq.(\ref{sym}), this holds true in any other frame. 
Applying the preceding arguments to the primed transformations, we 
arrive at the conclusion that the corresponding generators ${P_0}{'}$ of 
time translations are positive in all Lorentz frames and satisfy 
${P_0}' \, \vac =0$. Hence $\expec$ is a vacuum state, as claimed. 

\noindent (d) Let us finally demonstrate that, in spite of possible
ambiguities involved in the definition of generators and the ensuing 
interpretation of $\expec$,  
this state definitely does not describe a thermal 
equilibrium situation. To this end we show that $\expec$  
does not satisfy the KMS condition for any finite temperature 
$\beta^{-1} > 0$. Although the argument is standard, we present it 
here for the sake of completeness. Let $P_0$ be the non-negative 
generator defined in the preceding step. Then, there holds for all 
operators of the form $F(g) = \int dx_0\ g(x_0) F(x_0)$ the equality 
\begin{equation}
F(g)\, \vac = (2\pi)^{1/2} {\tilde g}(P_0) F \, \vac 
\end{equation}
where ${\tilde g}$ is the Fourier transform of $g$. Hence 
$F(g)\, \vac = 0$ if ${\tilde g}$ has its support on the negative real 
axis. Now, if $\expec$ satisfies the KMS condition for some $\beta$, it 
follows that for $F_1, F_2 \in {\cal F}$ we can continue analytically 
the function $x_0 \mapsto \langle\, F_1\, (F_2^{\dag} \, F(g))(x_0)\, \rangle$ 
to a function analytic in the complex domain 
$\{ z \in {\Bbb C}:\, 0 < \hbox {\rm Im}z <  \beta \}$ whose boundary 
value at Im$z = \beta$ is given by $x_0 \mapsto \langle\, (F_2^{\dag} \,
F(g))(x_0) \, F_1\, \rangle$. 
Since there holds $\langle\, F_1\, (F_2^{\dag} \, F(g))(x_0) \, \rangle = 
\lvac\, F_1 \, F_2^{\dag}(x_0) \, F(g)(x_0) \, \vac = 0$ for all $x_0 \in 
{\Bbb R}$, we find that $\langle\, F_2^{\dag}\, F(g)\, F_1\, \rangle = 
\lvac\, F_2^{\dag}\, F(g)\, F_1 \, \vac = 0$. As $F_1, F_2$ are arbitrary, 
we are thus led to the conclusion that $F(g)=0$. 
By applying the same argument to $F^{\dag}(g)$, we obtain similarly 
$F^{\dag}(g)=0$ and hence $F({\bar g})=F^{\dag}(g)^{\dag}=0$. Hence, 
$F(g)=0$ for any $g$ 
whose Fourier transform vanishes in some neighborhood of the 
origin. Therefore, the operator function $x_0 \mapsto F(x_0)$ is a 
polynomial in $x_0$ which can only be constant because of 
$\langle\, F(x_0)^{\dag}F(x_0)\, \rangle = \langle\, F^{\dag}F\, \rangle $  
by time invariance 
of $\expec$. Thus, since we can exclude the case of trivial dynamics, 
i.e., $F(x_0)= F$ for all $F \in {\cal F}$ and $x_0 \in 
{\Bbb R}$, the assumption that $\expec$ satisfies the KMS condition 
for some $\beta$ leads to a contradiction. 

Let us summarize: The existence of generators of supersymmetry 
(supercharges) in an arbitrary spatially homogeneous state $\expec$ 
implies that this state is supersymmetric in the sense of relation 
(\ref{sym}) and that the Bose-Fermi superselection rule is unbroken 
in this state. But any state with these two properties is necessarily 
a vacuum state and does not satisfy the KMS condition for finite 
$\beta$. Thus supersymmetry is broken in all thermal equilibrium 
states, irrespective of whether they describe 
pure or mixed homogeneous phases. Moreover, 
generators of supersymmetry cannot be defined in such states. Hence, 
thermal effects induce an inevitable spontaneous collapse of 
supersymmetry. 

\section{Role of Supertrace}
In discussions of thermal properties of supersymmetric theories, 
one frequently encounters so-called superthermal ensembles, described 
by non-positive ``density matrices''. It has been pointed out by van Hove 
\cite{vH} that thermal averages in these ensembles, called supertraces 
$\s$ in the following, ought to be interpreted as weighted differences 
of the underlying bosonic and fermionic subensembles, 
\begin{equation}
\s = w_b\ \expec_b - w_f\ \expec_f. \label{str}
\end{equation}
Here $\expec_b$, $\expec_f$ are the 
corresponding physical states and $w_b$, $w_f$ are non-negative 
numbers. Whenever this decomposition is meaningful, one can normalize 
these numbers according to $w_b + w_f =1$. 

It is sometimes argued \cite{C} that the behavior of $\s$ under the 
action of the supersymmetry transformations $\del$, $\delb$ provides 
the appropriate test for the spontaneous breakdown of supersymmetries. 
If $s(\del(\ \cdot\ ))=s(\delb(\ \cdot\ ))=0$, supersymmetry is said to 
be unbroken, otherwise it is regarded as spontaneously broken. 
This interpretation has no convincing conceptual basis, however, and it is 
therefore of some interest to explore the actual physical 
meaning of the two cases. Again the issue becomes very clear in 
the thermodynamic limit. It turns out that there exist only 
the following two possibilities in a spatially homogeneous situation: 
Either $\s=0$ (i.e., the supertrace is trivial) or the functionals
$s(\del(\ \cdot\ ))$ and $s(\delb(\ \cdot\ ))$ are different from 
$0$. Thus, superthermal ensembles can be supersymmtric only if they are 
trivial (having a zero ``density matrix''). 

Before going into the proof of this statement, let us briefly discuss 
its physical meaning. If $\s=0$, then $w_b = w_f$ and 
$\expec_b = \expec_f$, i.e., one cannot distinguish between a 
``bosonic'' and a ``fermionic'' phase. We emphasize that this does 
not imply the existence of only a single phase, because the state 
$\expec_b = \expec_f$ may well describe a mixture of different phases. 
On the other hand, if $\s \neq 0$, there are two possibilities; either 
(i) there are at least two different phases or (ii) there holds 
$\expec_b = \expec_f$ and $w_b \neq w_f$. It has been argued in 
\cite{vH} that the latter case does not occur in situations of physical 
interest. Yet since this argument is based on the existence of 
supercharges it is not applicable in the thermodynamic limit. In order 
to establish in this way the existence of different phases, therefore, 
one has to carry out further tests on $\s$. (It would be sufficient, 
for instance, to show that the superaverages of positive operators, 
$s(F^{\dag}F)$, can attain positive and negative values for suitable 
choices of $F \in {\cal F}$.)  

Thus, the supertrace may be used to 
obtain information about the phase structure of supersymmetric 
theories. Apart from the trivial case $\s = 0$, however, there is no 
restoration of supersymmetry at finite temperature even in the sense of 
the supertrace. 

We prove the above statement on $\s$ in two steps. First, we assume 
that both the bosonic and fermionic subensembles are pure phases. Then,  
as was explained in Sec.1, they have the cluster property, which we 
use here in a somewhat weaker form 
\begin{equation}
{1 \over \vert V \vert} \int_V d^3{\bold{x}}\ \langle\, F_1({\bold{x}}) 
F_2 \, \rangle_{b, f} - \langle \, F_1 \, \rangle_{b, f} \langle\, F_2 \, 
\rangle_{b, f}\ \ \rightarrow \ \ 0 
\qquad \hbox{as }V \nearrow {\Bbb R}^3. \label{clstr}
\end{equation}
If $\s$ is invariant under supersymmetry transformations, 
$s(\del(\ \cdot\ ))=0$, we obtain from the graded Leibniz rule 
(\ref{anti1})  
\begin{equation}
0 = s(\del (F_{-}({\bold{x}}) \, F_{+})) = s(\del (F_{-}({\bold{x}})) \,
F_{+} - F_{-}({\bold{x}}) \, \del (F_{+})) 
\end{equation}
for any $F_{\pm} \in {\cal F}_{\pm}$. Because of the decomposition  
(\ref{str}) of $\s$, this equality can be rewritten in the form 
\begin{equation}
w_b\ \langle \del (F_{-}({\bold{x}}))\, F_{+} - F_{-}({\bold{x}}) \, \del 
(F_{+}) \rangle_b = w_f\ \langle \del (F_{-}({\bold{x}})) \, F_{+} - 
F_{-}({\bold{x}}) \, \del (F_{+}) \rangle_f. \label{sub}
\end{equation}
Bearing in mind that $\del(\ \cdot\ )$ commutes with spatial 
translations, we thus obtain, by taking a spatial mean on both sides of 
(\ref{sub}) and making use of the cluster property (\ref{clstr}), 
\begin{equation}
w_b\ \big( \langle \del (F_{-}) \rangle_b \, \langle F_{+} \rangle_b - 
\langle F_{-} \rangle_b \, \langle \del (F_{+}) \rangle_b \big)
= w_f\ \big( \langle \del (F_{-}) \rangle_f \, 
\langle F_{+}\rangle_f - \langle F_{-}\rangle_f \, \langle \del (F_{+}) 
\rangle_f \big). \label{42}
\end{equation}
{}From the Bose-Fermi superselection rule (\ref{B-F}) applied to 
$\expec_{b,f}$, we get 
$\state{F_{-}}_{b, f} = 0$ and hence (\ref{42}) reduces to 
\begin{equation}
w_b \ \langle \del (F_{-})\rangle_b \, \langle F_{+} \rangle_b 
= w_f \ \langle \del (F_{-})\rangle_f \, \langle F_{+} \rangle_f. \label{sub1}
\end{equation}
The normalization condition 
$\langle {\bold{1}} \rangle_b = \langle {\bold{1}} \rangle_f = 1$ 
implies $w_b\ {\state{\del(F_{-})}}_b = w_f\ 
{\state{\del(F_{-})}}_f$; moreover, since $\expec_{b, \,  f}$ are 
thermal states, 
there exist, by the results obtained in Sec.2, operators $F_{-} \in 
{\cF}_{-}$ for which $\langle \del (F_{-})\rangle_{b, \, f} \neq 0$. 
Therefore, 
relation (\ref{sub1}) implies that $\expec_b = \expec_f$
and $w_b = w_f$ and hence we arrive at 
$\s = 0 $. 

Next, we discuss the general case where $\expec_{b, \, f}$ describe 
mixtures of thermodynamic phases. Then, we decompose the state 
$\expec = w_b\, \expec_{b} + w_f\, \expec_{f}$ into pure phases 
$\expec_{\theta}$ (central decomposition) as described in Sec.1, 
and get in particular 
\begin{eqnarray}
w_b\ \expec_b &=& \sum_{\theta} w_b(\theta)\ \expec_{\theta}, \label{bose_c}\\
w_f\ \expec_f &=& \sum_{\theta} w_f(\theta)\ \expec_{\theta}, \label{fermi_c}
\end{eqnarray}
where $w_b(\theta)$ and $w_f(\theta)$ are, respectively, non-negative 
weight factors and  $\sum_{\theta} w_b(\theta) = w_b, \  
\sum_{\theta} w_f(\theta) = w_f$. 

As in the case of pure phases, we proceed from the assumption of 
$s(\del(\ \cdot \ ))=0$ to relation (\ref{sub}), where we now insert 
the decompositions (\ref{bose_c}), (\ref{fermi_c}) of the bosonic and 
fermionic subensembles. Taking a spatial mean of the resulting 
expression and proceeding to the limit $V \nearrow {\Bbb R}^3$, we 
obtain, by applying the cluster property to each component pure phase 
$\expec_{\theta}$, the relation 
\begin{equation}
\sum_{\theta} w_b(\theta)\ 
\state{\del(F_{-})}_{\theta} \, \state{F_{+}}_{\theta}= 
\sum_{\theta} w_f(\theta)\ \state{\del(F_{-})}_{\theta} \,
\state{F_{+}}_{\theta}. \label{mix}
\end{equation}
To be precise, the interchange of the limit $V \nearrow {\Bbb R}^3$ with the 
summation $\sum_{\theta}$ requires some justification in the cases of 
an infinite number of phases, or of a continuum of phases (where, 
instead of the summation, an integration appears with a suitable 
probability measure). We refrain from presenting these technical 
details here. 

If one replaces in (\ref{mix}) the operator $F_{+}$ by 
$(1 / {\vert V \vert}) \int_V d^3{\bold{x}}\ 
(\del (F_{-})^{\dag})({\bold{x}}) \, F_{+}$ 
and makes use again of the cluster properties of pure phases, one 
obtains in the limit $V \nearrow {\Bbb R}^3$ the relation 
\begin{equation}
\sum_{\theta} w_b(\theta)\ \vert \state{\del(F_{-})}_{\theta} \vert^2
\, \state{F_{+}}_{\theta}= 
\sum_{\theta} w_f(\theta)\ \vert \state{\del(F_{-})}_{\theta} \vert^2 
\, \state{F_{+}}_{\theta}. \label{mixs}
\end{equation}
Repeating this procedure, one arrives at similar relations involving 
higher products of expectation values of 
arbitrary operators $F_+$ in the pure 
phases $\expec_{\theta}$. The resulting constraints on the weight 
factors $w_b(\theta), w_f(\theta)$ become obvious if one interprets 
(\ref{mixs}) as a relation between (non-normalized) states. Keeping 
$F_{-}$ fixed and varying $F_{+}$, one sees that the two functionals 
$\sum_{\theta} w_{b, \, f}(\theta)\ \vert \state{\del(F_{-})}_{\theta} 
\vert^2 \, \expec_{\theta}$ coincide. Because of the uniqueness of 
the central decomposition this implies for any $\theta$ the equality 
\begin{equation}
w_b(\theta)\ \vert \state{\del(F_{-})}_{\theta} \vert^2 = w_f(\theta)\ \vert 
\state{\del(F_{-})}_{\theta} \vert^2. 
\end{equation}
Since $\expec_{\theta}$ is a thermal state, however, there is some 
$F_{-} \in \cF_{-}$ such that $\state{\del(F_{-})}_{\theta} \neq 0$ and 
hence we obtain $w_b({\theta}) = w_f({\theta})$. It then follows from 
relations (\ref{bose_c}), (\ref{fermi_c}) that $\s = 0$, so only the 
trivial supertrace is supersymmetric. 

Thus we conclude that the supertrace is a device to deduce some 
(partial) information about the phase structure in supersymmetric 
theories. Yet its behaviour under supersymmetry transformations does 
not provide any additional information, in accord with the result of 
the previous section that supersymmetry always suffers from a spontaneous 
collapse in thermal states. 

\section{Conclusions}
\noindent In the present article, we have clarified in a general 
setting the status of supersymmetry in thermal states. In every quantum 
field theory where an action of supersymmetry transformations on the 
fields can be so defined that the fundamental relation (\ref{susy}) 
holds, this symmetry suffers from a spontaneous collapse in thermal 
states. We have established this result for spatially homogeneous 
states in d=4 dimensions; but it can easily be extended to more complex 
situations (such as asymptotically homogeneous states, spatially 
periodic states, etc.) and to any number of spacetime dimensions. Moreover, 
the point-like nature 
and strict local (anti-) commutativity of the underlying fields is 
not really crucial. Our arguments require only a sufficiently rapid 
fall-off of the expectation values of (anti-) commutators of the 
underlying field operators for large spatial translations. 
Therefore, an analogous result may be 
expected to hold in quantum superstring field theory, provided a pertinent 
formulation of supersymmetry can be given in that setting.  

The universal breakdown of certain symmetries in thermal states is a well 
known phenomenon. A prominent example is the Lorentz symmetry which  
is inevitably broken in thermal equilibrium states, since the KMS 
condition fixes a rest frame \cite{LorSSB}. Nevertheless, an action of 
Lorentz transformations can be defined on thermal states and is 
physically meaningful: A gas which is macroscopically at rest in a 
given Lorentz frame is transformed into a gas in motion with respect to 
that frame, etc.\,(cf.\ \cite{BB} for a general characterization of 
thermal equilibrium states in arbitrary Lorentz frames). 

This familiar situation of spontaneous breakdown of a symmetry should 
be clearly distinguished from the spontaneous collapse of supersymmetry 
in thermal states, where it is no longer possible to define an action 
of the symmetry on the physical states. 

In view of this vulnerability 
to thermal effects, one may wonder how supersymmetry can manifest 
itself in real physical systems. The theoretical prediction of a zero 
energy mode in thermal states is of limited value, since this 
mode need not be affiliated with a Goldstino particle, but may result 
from long range correlations between particle-hole pairs \cite{MPU}.  
Also, rigorous results on the fate of particle 
supermultiplets in a thermal environment do not exist yet. 

For a reliable prediction of the existence of supersymmetry in physical 
systems, it seems necessary to show that symmetry properties of 
the vacuum theory can be \mbox{recovered} from thermal states in the 
limit of 
zero temperature. On the other hand, the possibility that supersymmetry 
remains collapsed in this limit, in analogy to some hysteresis effect, 
may be even more interesting since it could account for the apparent 
absence of this symmetry in the real world. It would therefore be 
desirable to clarify which of these two possibilities is at hand in 
models of physical interest.

\vskip15pt
\noindent {\bf \Large Acknowledgements}\\[1mm] 
\noindent One of the authors (D.B.) gratefully acknowledges the 
hospitality extended to him at RIMS, Kyoto University, and financial 
support from the Japanese Society for the Promotion of Science and 
the Alexander von Humboldt Foundation. Both authors would like to 
thank Prof.\ T.\ Kugo for an exposition of the present status of 
supersymmetry in quantum superstring field theory.


\begin{thebibliography}{99}
\bibitem{A} L.\  Girardello, M.\ Grisaru and P.\ Salomonson, Nucl.\ Phys.\ 
{\bf B178} (1981) 331; \\
D.\ Boyanovsky, Phys.\ Rev.\ {\bf D29} (1984) 743. 
%
\bibitem{B} J.\ Fuchs, Nucl.\ Phys.\ {\bf B246} (1984) 279; \\ 
Won-Ho Kye, Sin Kyu Kang and Jae Kwan Kan, Phys.\ Rev.\ {\bf D46} (1992) 1835
%
\bibitem{C} L.\ van Hove, Nucl.\ Phys.\ {\bf B207} (1982) 15; \\
T.E.\ Clark and S.T. Love, Nucl.\ Phys.\ {\bf B217} (1983) 349. 
%
%
\bibitem{HHW} R.\ Haag, N.M.\ Hugenholtz and M.\ Winnink, 
Comm.\ Math.\ Phys.\  
{\bf 5} (1967) 215; \\ I.\ Ojima, 
Ann.\ Phys.\ {\bf 137} (1981) 1. 
%
\bibitem{BR} O.\ Bratteli and D.W.\ Robinson, Operator Algebras 
and Quantum Statistical Mechanics, Vols.\ 1 and 2 (Springer, Berlin, 1979 and 
1981). 
%
\bibitem{MTU} H.\ Matsumoto, M.\ Tachiki and H.\ Umezawa, 
Thermo Field Dynamics and Condensed States (North-Holland, Amsterdam, 1982). 
%
\bibitem{GNS} R.F.\ Streater and A.S.\ Wightman, 
PCT, Spin and Statistics and All That (Benjamin, New York, 1964); \\ R.\ Haag, 
Local Quantum Physics (Springer, Berlin, 1992[1st ed.], 1996[2nd ed.]). 
%
\bibitem{PTW} M.B.\ Paranjape, A.\ Taormina and L.C.R.\ Wijewardhana, Phys.\ 
Rev.\ Lett.\ {\bf 50} (1983) 1350. 
%
\bibitem{Reeh} H.\ Reeh, Lecture Notes in Physics {\bf 39} 
(H.\ Araki, ed.) (Springer, Berlin, 1975) pp.249-255. 
%
\bibitem{vH}L.\ van Hove, Fizika {\bf 17} (1985) 267. 
%
\bibitem{LorSSB} I.\ Ojima, Lecture Notes in Physics {\bf 176} (K.\ Kikkawa, 
et al.\ eds.) (Springer, Berlin, 1983) pp.161-165; 
Lett.\ Math.\ Phys.\ {\bf 11} (1986) 73. 
%
\bibitem{BB} J.\ Bros and D.\ Buchholz, Nucl.\ Phys.\  {\bf B429} (1994) 291.
%
\bibitem{MPU} H.\ Matsumoto, N.J.\ Papastamatiou, H.\ Umezawa and N.\ 
Yamamoto, Phys.\ Rev.\ {\bf D34} (1986) 3217. 
\end{thebibliography}
\end{document}